# Overcoming the Fundamental Barrier Thickness Limits of Ferroelectric Tunnel Junctions through BaTiO$_3$/SrTiO$_3$ Composite Barriers

*Lingfei Wang,*[†,‡] *Myung Rae Cho,*[†,‡] *Yeong Jae Shin,*[†,‡] *Jeong Rae Kim,*[†,‡] *Saikat Das,*[†,‡] *Jong-Gul Yoon,*[⊥] *Jin-Seok Chung*[§] *and Tae Won Noh*[*,†,‡]

[†]Center for Correlated Electron Systems, Institute for Basic Science (IBS), Seoul 08826, Republic of Korea.

[‡]Department of Physics and Astronomy, Seoul National University, Seoul 08826, Republic of Korea.

[⊥]Department of Physics, University of Suwon, Hwaseong, Gyunggi-do 445-743, Republic of Korea.

[§]Department of Physics, Soongsil University, Seoul 156-743, Republic of Korea.

ABSTRACT:

Ferroelectric tunnel junctions (FTJs) have attracted increasing research interest as a promising candidate for non-volatile memories. Recently, significant enhancements of tunneling electroresistance (TER) have been realized through modifications of electrode materials. However, direct control of the FTJ performance through modifying the tunneling barrier has not been adequately explored. Here, adding a new direction to FTJ research, we fabricated FTJs with $BaTiO_3$ single barriers (SB-FTJs) and $BaTiO_3/SrTiO_3$ composite barriers (CB-FTJs), and reported a systematic study of FTJ performances by varying the barrier thicknesses and compositions. For the SB-FTJs, the TER is limited by pronounced leakage current for ultrathin barriers and extremely small tunneling current for thick barriers. For the CB-FTJs, the extra $SrTiO_3$ barrier provides an additional degree of freedom to modulate the barrier potential and tunneling behavior. The resultant high tunability can be utilized to overcome the barrier thickness limits and enhance the overall CB-FTJ performances beyond those of SB-FTJ. Our results reveal a new paradigm to manipulate the FTJs through designing multilayer tunneling barriers with hybrid functionalities.

Ferroelectric tunnel junctions (FTJs), composed of a thin ferroelectric (FE) layer sandwiched in between two metallic electrodes, have been intensively investigated in recent years. This device is considered to be a promising candidate for next-generation nonvolatile memories, because it combines the advantages of both ferroelectric random-access-memory and resistive-switching memory.[1-3] The concept of FTJ was first proposed by Esaki in 1971,[4] but the research activities have not flourished untixl this decade.[5-21] The operation of FTJs has been mainly explained in terms of interfacial screening of polarization charges.[5-8] In a metal 1(M1)/FE/metal 2 (M2) structured FTJ (Figure 1a), the two asymmetric electrodes lead to unequal screen lengths and potential changes at metal/FE interfaces. Depending on the polarizations, the electrostatic potential profile of tunneling barrier will be varied. As a result, the tunneling resistance can be switched between low (ON) and high (OFF) values by polarization reversal, leading to the so-called tunneling electroresistance (TER) effect.

Extensive experimental works based on this asymmetric screening scenario have been reported during the last decade. In 2009, the TER effect was first demonstrated using conducting atomic force microscopy (CAFM).[9-11] Subsequently, FTJs with highly reproducible performance were realized in capacitor geometry, which will be useful for practical applications.[12,13] Up to recently, most studies of FTJs have focused on improving device performance by modifying the electrode materials.[14-20] Typical examples include the use of lightly doped semiconductors,[14-16] correlated electron oxides with metal-insulator transitions,[16-18] graphene/molecular bilayers,[19] and high work function metals[20] as electrodes.

Nevertheless, for any tunneling device, modulating the tunneling barrier should be a more straightforward approach to controlling the device performance. For a simple M1/FE/M2 FTJ, Zhuravlev *et al.* predicted that the TER is sensitive to both the polarization and the barrier thickness.[5] It has been demonstrated that FE barrier with a higher polarization (such as "super-tetragonal" or doped BiFeO$_3$) can significantly boost the TER effect.[15,21] On the other hand, experimentally varying the FE barrier thickness is accompanied by additional complicated changes. An ultrathin barrier results in a large tunneling current, which facilitates easy signal readout and device scaling. However, serious undesirable effects, including the FE dead layer,[22] pinned interface dipole,[23-25] and leakage current,[11,18,26] arise and degrade the device performance. For a thick barrier, these obstacles can be avoided, but the tunneling current becomes too small for realizing a practically useful device. As an alternative, several groups have theoretically proposed a new kind of FTJ using an FE/paraelectric (FE/PE) composite barrier.[27,28] The PE layer provides a new route to control the tunneling barrier potentials, and thus the TER can be tuned and significantly enhanced. However, there have been few experimental efforts to systematically attest this theoretical prediction.[29]

In this paper, we report experimental control of the TER effect by directly manipulating the tunneling barrier thickness and composition. For this purpose, we fabricated two types of FTJs: BaTiO$_3$ (BTO) single-barrier FTJs (SB-FTJs) and BaTiO$_3$/SrTiO$_3$ (BTO/STO) composite-barrier FTJs (CB-FTJs). As schematically depicted in Figures 1a and b, the electrostatic potential profiles of these two types of FTJs are distinct. For the SB-FTJs, thickness is the only tunable parameter of the tunneling barrier. As a result, the device performance is fundamentally limited by pronounced leakage current for ultrathin barriers and by extremely small tunneling current for thick barriers. On the other hand, for the CB-FTJ, the extra STO layer provides an additional degree of freedom for tuning the tunneling behavior. The resultant high tunability makes the CB-FTJ a feasible device structure for circumventing the fundamental barrier thickness limits of SB-FTJ. By quantitatively comparing the enhanced TER with a simple tunneling model, we will show that the work mechanism of our CB-FTJs can be understood in terms of the barrier-composition-modulated electrostatic potential profile.

Figures 1c and d show the schematic device architectures of an SB-FTJ [Au/Ti/BTO/SrRuO$_3$/STO(001)] and a CB-FTJ [Au/Ti/BTO/STO/SrRuO$_3$/STO(001)], respectively. The oxide layers were grown by pulsed laser deposition on atomically smooth TiO$_2$-terminated STO(001) substrates. We first deposited a 20 nm-thick SrRuO$_3$ (SRO) film as the bottom electrode layer, and then deposited BTO and STO layers of various thicknesses ($t_{BTO}$ and $t_{STO}$). The BTO and STO film thicknesses were monitored by high pressure reflection high-energy electron diffraction (RHEED) intensity oscillations (see Methods section, Figures S1 and S2 in Supporting Information for details). For the SB-FTJs, we varied $t_{BTO}$ from 2 to 10 uc. For the CB-FTJs, we varied $t_{BTO}$, but fixed the total barrier thickness ($t_{total} = t_{BTO} + t_{STO}$) at 10 uc. The top electrodes were sequentially deposited with 5 nm Ti and 25 nm Au films. Using electron-beam lithography and one-step lift-off techniques, we obtained circular top electrodes of 500 nm in diameter (inset of Figure 2c). Such small-sized top electrodes were required to minimize the leakage current. During electrical measurements, the top electrode was grounded through a CAFM tip, and the voltage bias was applied through the bottom SRO electrode (Figure 1c).

All of the SB- and CB-FTJs are composed of high-quality epitaxial oxide films. Atomic force microscopy (AFM) topographic images (Supporting Information Figures S3 and S4) display uniform unit-cell-height terraces. X-ray diffraction (XRD) $\omega$-$2\theta$ linear scans show clear Laue fringes (Supporting Information Figure S5), which further confirm good epitaxial film qualities. We estimated the lattice constants of the oxide layers from XRD reciprocal space mapping (RSM) (Supporting Information Figure S6). For the 10 uc BTO films, the in-plane and out-of-plane lattice constants were 3.95 and 4.21 Å, respectively. Such a large tetragonality of 1.07 signifies a robust FE polarization in these BTO thin films.[30,31] Piezoresponse force microscopy (PFM) measurements (Supporting Information Figures S7 and S8) further confirmed the FE hysteresis loops for all the BTO and BTO/STO films.

Our FTJ devices exhibit highly reproducible electrical properties. As an example, Figure 2 shows the performance of a SB-FTJ device with $t_{BTO}$ = 8 uc at room temperature. After a 100 ms voltage pulse with $V_{write}$ = +8.0 V (-8.0 V) was applied, the high resistance OFF state (low resistance ON state) was obtained. Figure 2a shows the nonlinear current-voltage (*I-V*) curves between ±1.5 V in the ON and OFF states. The ON state current ($I_{ON}$) is much larger than that of the OFF state ($I_{OFF}$). As shown in Figure 2b, we calculated TER = ($I_{ON}$ - $I_{OFF}$)/$I_{OFF}$. The voltage-dependent TER shows a maximum of 53,000% around ±1.0 V (optimal voltage). Figures 2c and d show the *I-V* curves and ON/OFF resistance states at a fixed voltage bias ($V_{read}$) of -1.0 V from 50 devices, respectively. The data exhibit little fluctuation, demonstrating the uniformity and reliability of our FTJ devices. We also performed endurance test up to 2,000 cycles and found that the TER remained nearly the same (Supporting Information Figure S10).

The observed resistance change between the ON and OFF states should originate from FE polarization switching. Note that extrinsic mechanisms for FTJs have also been proposed, including the interfacial charge redistribution and electrochemical reactions.[32,33] To rule out such extrinsic effects, we need to show the direct relationship between FE and TER effect. We initially poled the SB-FTJ device ($t_{BTO}$ = 8 uc) and applied a 100 ms $V_{write}$ pulse, then measured the resistance (*R*) at $V_{read}$ = -1.0 V (see Supporting Information Figures S11a-d for details). By varying the poling direction and $V_{write}$ value, we could obtain *R*-$V_{write}$ curves (Figure 2e), which exhibit well-defined resistive-switching behaviors near the coercivity of PFM loops (Figure 2f). The corresponding value was about 51,000%, in good agreement with that estimated from the *I-*

*V* curve in Figure 2a. These results demonstrate that the TER is indeed dominated by FE polarization-modulated tunneling phenomena.[34]

Before describing our CB-FTJ work, we will address two fundamental barrier thickness limits of SB-FTJ performances. Given an ultrathin FE barrier, the SB-FTJ performance will be limited by pronounced leakage current. Figures 3a-c shows the scanning probe microscopy results for the BTO/SRO/STO(001) heterostructure with $t_{BTO}$ = 2 uc. First, we wrote antiparallel FE domains by applying ±8.0 V biases. The out-of-plane PFM phase images (Figure 3a) clearly show contrasts of 180 degrees for the antiparallel FE domains. Figure 3b shows the CAFM current mappings. The domains with upward polarization (pointing toward the surface) exhibit a much smaller tunneling current than that of the downward polarized domains, signifying the existence of the TER effect.[10,11] However, the curly pattern in the CAFM image clearly shows highly non-uniform current distribution within an FE domain. We measured the current and corresponding topology profiles along the red arrow in Figure 3b. As shown in Figure 3c, the local currents through the terrace edge are indeed several times larger than those within the terrace. This contrast can be explained by the sample-CAFM tip contact area change and the large leakage current induced by defects near terrace edges.[35] We suggest that the later effect, defect-mediated leakage current, should play the dominating role on the local current variation in our ultrathin BTO sample (see Supporting Information, Section 4 for detailed discussions). Because such structural imperfections at terrace edges are unavoidable, the leakage current could pose a fundamental obstacle to signal readout of SB-FTJs.

The leakage current is also evident in the *I-V* curves of our ultrathin SB-FTJs (Supporting Information Figure S12). For SB-FTJs with $t_{BTO} \leq 3$ uc (Supporting Information Figure S12a), the leakage current dominates the transport behavior, such that both $I_{ON}$-*V* and $I_{OFF}$-*V* curves are linear and almost identical. With increasing $t_{BTO}$ above 4 uc (Supporting Information Figures S12b-f), the $I_{ON}$-$V$ and $I_{OFF}$-$V$ curves become more non-linear and show meaningful differences. The CAFM images of BTO films with $t_{BTO}$ = 4 and 6 uc also consistently show strongly suppressed local conductivity variations at terrace edges (Supporting Information Figure S9). These observations suggest that the leakage current only crucially affects the tunneling behavior of SB-FTJ with ultrathin barriers.

For SB-FTJs with thicker barriers, the tunneling current can decrease below the experimental detection limit. According to the Wentzel-Kramers-Brillouin (WKB) theory of quantum mechanical tunneling,[36] the tunneling probability should decrease exponentially with increasing barrier thickness. Figures 3d and e shows $I$-$V$ curves on a logarithmic scale for SB-FTJ with $t_{BTO}$ = 8 and 10 uc, respectively. In both cases, $I_{OFF}$ becomes orders of magnitude smaller than $I_{ON}$. Moreover, $I_{OFF}$ approaches a detection limit caused by the noise current background of our experimental setup, i.e., approximately 0.5 pA at room temperature. For $t_{BTO}$ = 10 uc, the true $I_{OFF}$ value within ±1 V is completely hidden below the noise current background.

These two fundamental barrier thickness limits of SB-FTJs appear clearly in the $t_{BTO}$-dependent tunneling currents. Figure 4a shows both $I_{ON}$ and $I_{OFF}$ values at -0.4 V. In the regime of $t_{BTO} \geq 6$ uc, both $I_{ON}$ and $I_{OFF}$ increase exponentially with decreasing $t_{BTO}$, consistent with typical tunneling behavior.[5,27] The dotted curves in Figure 4a show the theoretical predictions of $I_{ON}$ and $I_{OFF}$ based on Brinkerman's tunneling model,[10,16,37] which will be described later. As $t_{BTO}$ decreases below 5 uc, both $I_{ON}$ and $I_{OFF}$ start to exceed the dotted curves, due to the large leakage current contribution. On the other hand, with increasing $t_{BTO}$, $I_{OFF}$ decreases and reaches the experimental detection limit of ~0.5 pA. Specifically, in the case of the SB-FTJ with $t_{BTO}$ = 10 uc, the experimental $I_{OFF}$ value (marked by the open square symbol in Figure 4a) arises not from tunneling, but mainly from noise.

These fundamental barrier thickness limits will significantly degrade the performance of SB-FTJs. The TER values at -0.4 V were extracted from the *I-V* curves (Figure 4b). The dotted red line corresponds to the ideal TER values obtained from the fitting curves in Figure 4a. In the small-$t_{BTO}$ regime ($\leq$ 5 uc), the TER becomes much smaller than the ideal value. Obviously, this degradation is due to the large leakage current. In the large-$t_{BTO}$ regime (> 8 uc), the TER also becomes degraded, due to the measurement limit of small $I_{OFF}$. These two degradations result in a non-monotonic TER-$t_{BTO}$ dependence, and establish a bound for the maximum TER in SB-FTJs. The best performance of our SB-FTJ was observed at $t_{BTO}$ = 8 uc. The TER at $V_{read}$ = -0.4 V is 10,000%, and the value at the optimum $V_{read}$ = -1 V is 53,000% (Supporting Information Figure S13). Note that the two fundamental limits addressed above should be generic in any type of SB-FTJ device.[9,11,18,38]

We now look into the CB-FTJs, composed of the BTO/STO composite barriers. Simplified electrostatic potential profiles are depicted in Figure 1b. At the Ti/BTO interface, the polarization charges will be screened by the accumulated/depleted carriers inside the Ti metal. For the opposite side, the nonpolar STO provides far fewer screening charges at the BTO/STO interface. Hence, most screening occurs at the STO/SRO interface, and a large depolarization field is generated inside the STO layer.[28] As theoretically noted in the literatures[27,28], due to the depolarization field, this STO layer can serve as a switch that changes its barrier height from a low to a high value when the BTO polarization direction is reversed. This change in electrostatic potential will affect the TER substantially.

In the CB-FTJ structure, we can control the tunneling current by varying the barrier composition (i.e., varying $t_{BTO}$ and $t_{STO}$). For simplicity, we fixed $t_{total}$ at 10 uc to avoid the undesirable leakage current, then tuned $t_{BTO}$ from 2 to 8 uc (the corresponding $t_{STO}$ varies from 8 to 2 uc). The *I-V* curves are presented in Figures 5a-d (linear scale) and in Supporting Information Figure S14 (logarithmic scale). Figure 6a shows the corresponding values of both $I_{ON}$ and $I_{OFF}$ plotted using red open circles and black open squares, respectively. $I_{OFF}$ continues to increase with decreasing $t_{BTO}$. In contrast, $I_{ON}$ initially increases but then decreases below $t_{BTO}$ = 6 uc. The complicated behavior of $I_{ON}$ actually suggests that the barrier potential profiles do evolve with $t_{BTO}$, which will be addressed later.

We could achieve considerable TER enhancement using the CB-FTJ structure. Figure 6b shows TER values for our CB-FTJs with various BTO/STO barriers. The best performance was obtained at $t_{BTO}$ = 6 uc. The TER calculated at $V_{read}$ = -0.4 V is 36,000%, more than 10 times higher than that of the SB-FTJ with the same total barrier thickness (2,600% at $t_{total} = t_{BTO} = 10$ uc). The TER at optimum $V_{read}$ = 1.1 V is as high as 170,000% (Supporting Information Figure S13), which is even higher than that of the best SB-FTJ (i.e., 53,000% at $t_{BTO}$= 8 uc). In addition, for the CB-FTJ with $t_{BTO}$ = 4 uc (2 uc), the optimal TER can reach 9,200% (140%), much higher that of SB-FTJs with the same $t_{BTO}$. This result further confirms that the additional STO layer can effectively reduce the leakage current.

Figure 5e shows the hysteretic resistive-switching behavior of this best performance sample ($t_{BTO}$ = 6 uc) in terms of $V_{write}$ (The resistance is measured at optimal $V_{read}$ = 1.1 V; see Supporting Information Figures S11e and f for details). The switching voltage is also consistent with the coercivity of the PFM hysteresis loop (Figure 5f), confirming the polarization-dominated operation of CB-FTJs. The TER ratio calculated from $R_{ON}$ and $R_{OFF}$ at $V_{write}$ = 0 V is

~160,000%, in good agreement with the value estimated from corresponding *I-V* curves. To our knowledge, this optimal TER is higher than those of reported FTJs using a single BTO tunnel barrier and conventional metal electrodes.[3,10,11,38]

The high tunability and enhanced TER can be quantitatively understood using a quantum mechanical tunneling model with various electrostatic potential barriers. First, we applied Brinkerman's tunneling model to the SB-FTJs. We assumed that the tunneling barrier potential profile above the Fermi level has a trapezoidal shape (Figure 1a), and calculated the tunneling current based on the WKB approximation (see Supporting Information Section 6 for details).[37] As shown in Supporting Information Figures S16a-f, we can fit the *I-V* curves of SB-FTJs in the small bias regime (±0.4 V) quite well. The fitting parameters of the barrier potential are summarized in Supporting Information Table S1, and the real space potential profiles of the ON [$j_{ON}(x)$] and OFF [$j_{OFF}(x)$] states are shown in Figures 6c-g. At $t_{BTO} \geq 6$ uc, the barrier potentials are highly asymmetric and show distinct heights between the ON and OFF states. At $t_{BTO} < 6$ uc, however, the barrier potential asymmetry and height difference between $j_{ON}$ and $j_{OFF}$ decrease significantly, signifying a degradation of FE polarization. Note that the potential height could also be underestimated due to the large leakage current. In addition, we further simplified Brinkerman's model by treating the barrier as a rectangular potential with unequal heights for the ON ($\bar{\varphi}_{ON}$) and OFF ($\bar{\varphi}_{OFF}$) states (average-barrier-approximation, see Supporting Information Section 6).[10,16] The calculated $I_{ON}$ ($I_{OFF}$)-$t_{BTO}$ and TER-$t_{BTO}$ curves from this tunneling model are plotted in Figures 4a and b, respectively.

We then considered the barrier potential modulations in CB-FTJs. As described in Figure 1b and Supporting Information Figure S17, the potential of STO is flat, and its height is pinned to that of BTO at the BTO/STO interface.[27,28] Thus, we can treat the potential profile of the BTO/STO composite barrier as a stack consisting of a trapezoidal barrier (i.e., BTO) and a rectangular barrier (i.e., STO). For simplicity, we assumed that the BTO potential profiles remain the same as those of the SB-FTJ case with the same $t_{BTO}$. We define the STO barrier height difference between the ON and OFF states as $\Delta\varphi_{STO}$. The schematic barrier potential profiles of CB-FTJs with various $t_{BTO}$ are also plotted in Figures 6c-g.

In the CB-FTJ performance, $\Delta j_{STO}$ and the depolarization field are competing against each other. A thicker STO layer gives rise to a larger contribution of $\Delta j_{STO}$ to the barrier potential profile. Namely, it magnifies the difference between $j_{ON}$ and $j_{OFF}$. In the regime of $t_{BTO} \geq 6$ uc, this effect should dominate the TER of the CB-FTJ. On the other hand, when $t_{BTO}$ is reduced below 4 uc, there will be a considerable degradation of FE polarization because of the large depolarization field.[31,39,40] As a result, the ON/OFF difference of the BTO barrier potential inevitably decreases substantially. The competition between $\Delta j_{STO}$ and the depolarization field results in the non-monotonic $t_{BTO}$-dependence of the tunneling current, which can tune TER in the CB-FTJs.

For the CB-FTJ with $t_{BTO} = 6$ uc, $I_{ON}$ and $I_{OFF}$ are well tuned in a proper range to overcome the fundamental barrier thickness limits, i.e. due to the leakage current and the experimental current detection. We quantitatively evaluated this TER enhancement based on theoretical calculations (see Supporting Information, Section 7 for details). As shown in Supporting Information Table S2, we first calculated the average barrier potential height. $\bar{\varphi}_{OFF}$ (0.72 eV) is comparable with those of the SB-FTJs with $t_{BTO} \geq 8$ uc ($\geq 0.76$ eV), whereas $\bar{\varphi}_{ON}$ (0.29 eV) is much smaller than those of the SB-FTJs (0.41 eV for $t_{BTO} = 8$ uc and 0.45 eV for $t_{BTO} = 10$ uc). Accordingly, the increase of $I_{ON}$ is responsible for the enhanced TER beyond the maximum value in SB-FTJs. We then calculated the TER value of this sample based on the WKB approximation. The calculated

TER at -0.4 V is 56,000%, which agrees reasonably well with our experimental value of 36,000%. This agreement further demonstrates that the large TER in CB-FTJ originates from the changes in the barrier potential profiles.

In summary, we experimentally realized atomic-scale control of TER in the CB-FTJ devices by incorporating BTO/STO composite barriers. As proved by the experimental results and theoretical calculations, the composite barriers enabled an effective control of the barrier potentials and tunneling currents. The high tunability can be utilized to circumvent two fundamental barrier thickness limits of TER in SB-FTJs: i) the pronounced leakage current through ultrathin barriers and ii) small tunneling current below the experimental detection limit for thick barriers. These results offer a new approach to modify and improve the FTJ performances by designing multilayer tunneling barriers with hybrid functionalities.

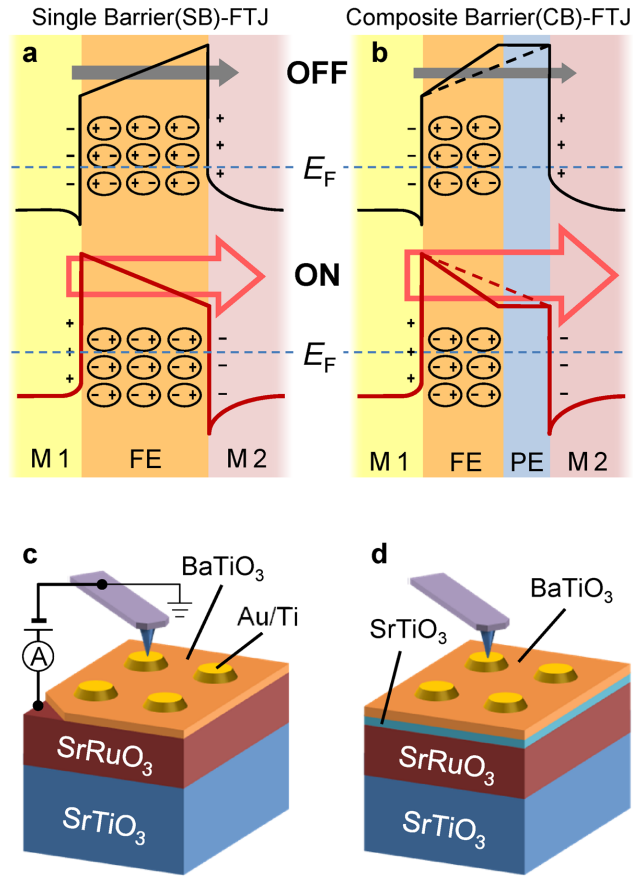

**Figure 1.** Schematic electrostatic potential profiles and device architectures of FTJs. (a,b) Schematic electrostatic potential profiles of (a) a SB-FTJ and (b) a CB-FTJ. For the CB-FTJ, a part of the FE barrier is replaced by a PE layer. The Fermi levels ($E_F$) are marked by dotted lines. The barrier potential of SB-FTJ above the $E_F$ approximately has a trapezoidal shape. The directions of FE polarization are illustrated by open ellipses, and the accumulated (depleted) space charges at metal/FE interfaces are illustrated by "+" ("−"). The solid (gray) and open (pink) arrows represent the magnitudes of tunneling currents in the OFF and ON states. Note that, in (b), the potential profiles of SB-FTJ are also plotted as dashed lines for comparison. For a fixed $t_{total}$, the CB-FTJ indeed has a lower (higher) potential height in the ON (OFF) state than that of the SB-FTJ. As a result, we expect a larger (smaller) $I_{ON}$ ($I_{OFF}$). (c,d) Schematic device architectures of (c) an SB-FTJ [Au/Ti/BTO/SRO/STO(001), BTO single barrier] and (d) a CB-FTJ

[Au/Ti/BTO/STO/SRO/STO(001), BTO/STO composite barrier]. As depicted in (c), during the electrical measurements the voltage bias is applied to the SRO bottom electrode, and the Au/Ti top electrode is grounded by a conductive CAFM tip.

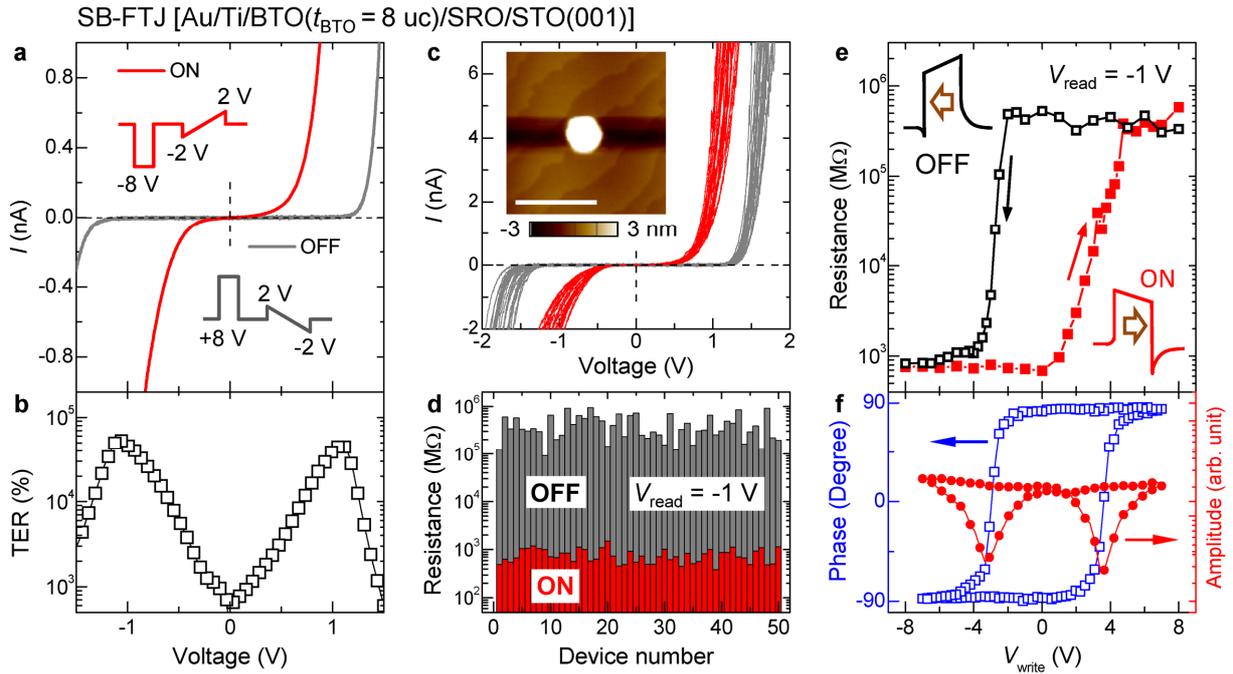

**Figure 2.** TER and resistive-switching behavior of the SB-FTJ with $t_{BTO}$ = 8 uc. (a) $I_{ON}$-$V$ and $I_{OFF}$-$V$ curves. The inset of (a) illustrates the applied voltage sequences for the $I$-$V$ measurements. (b) Voltage dependent TER calculated from the $I$-$V$ curves in (a). (c) $I$-$V$ curves measured from 50 different devices. The inset of (c) shows the AFM topographic image of one FTJ device with an Au/Ti top electrode. The scale bar corresponds to 1 μm, and the diameter of the Au/Ti top electrode is ~500 nm. (d) $R_{ON}$ and $R_{OFF}$ measured at $V_{read}$ = -1 V from 50 different devices. The results in (c) and (d) demonstrate that our FTJ devices are highly uniform and reproducible. (e) $R$-$V_{write}$ curves with $V_{read}$ = -1 V after successive voltage pulses of 100 ms were applied. (f) PFM phase-voltage hysteresis loop and amplitude-voltage butterfly loop. The data in (e) and (f) clearly show that the resistive-switching behavior occurs at the FE coercive field.

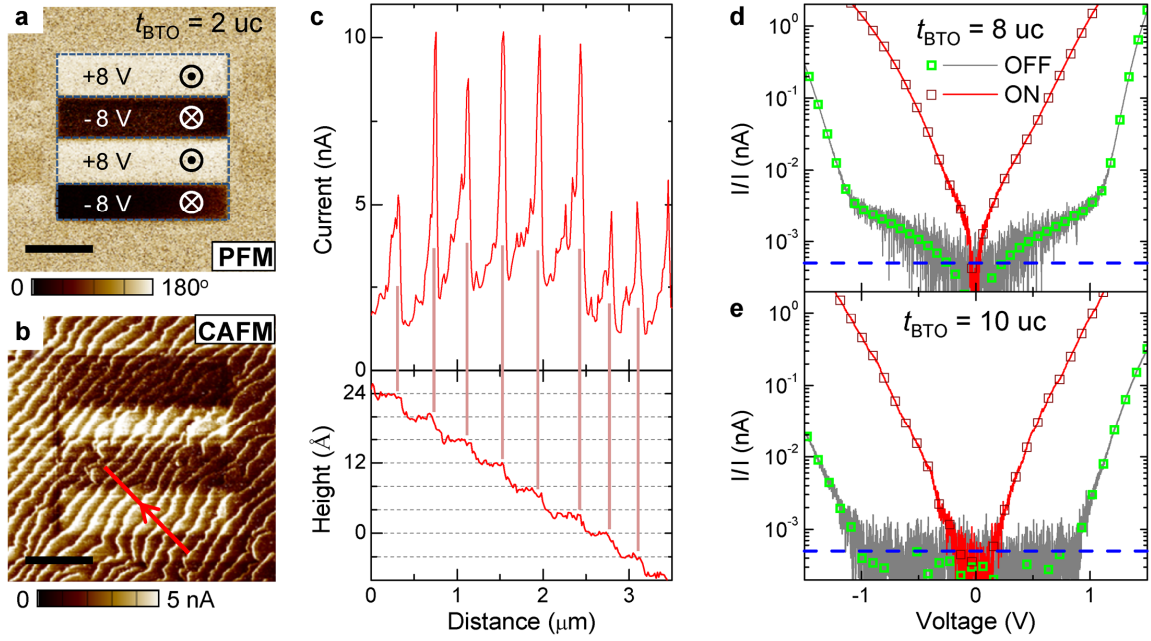

**Figure 3.** FTJ characterizations to demonstrate two barrier thickness limits of the device performances. (a) Out-of-plane PFM phase image and (b) CAFM current mapping image of the BTO/SRO/STO(001) heterostructure with $t_{BTO}$ = 2 uc. For the PFM image, antiparallel domains are written by ±8 V. The dark (bright) contrasts represent the downward (upward) polarization. During the current mapping via CAFM, a -1 V bias is applied to the bottom SRO electrode. The scale bars in all images correspond to 2 μm. (c) Current and topographic line profiles along the solid (red) arrow in (b). The vertical lines in (c) indicate that the leakage current is closely related to the terrace edges. (d,e) Logarithmic *I-V* curves of SB-FTJ with $t_{BTO}$ = (d) 8 uc and (e) 10 uc. The original experimental curves are plotted by solid lines (red or gray), and the averaged values are represented by open squares (wine or green). Note that $I_{OFF}$ of both samples reaches the detection limit caused by the noise background of our experimental setup. The noise current of approximately 0.5 pA is marked by the dashed lines in (d) and (e).

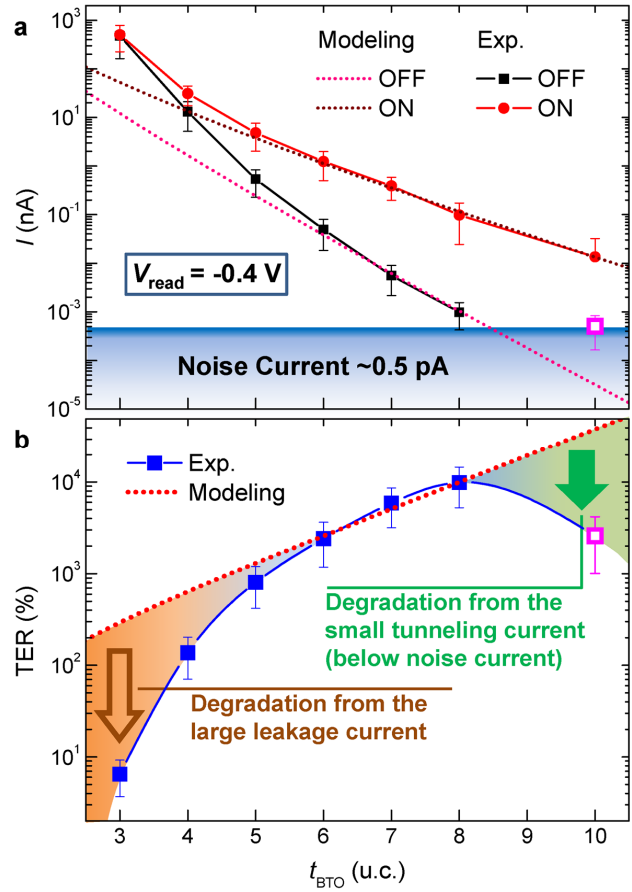

**Figure 4.** Barrier-thickness-dependent tunneling current and TER in SB-FTJ. (a) $t_{BTO}$ dependent $I_{ON}$ and $I_{OFF}$ at -0.4 V. The dotted lines represent the $I_{ON}$-$t_{BTO}$ and $I_{OFF}$-$t_{BTO}$ curves predicted by the Brinkerman's tunneling model. Our experimental detection limit estimated from the noise current is approximately 0.5 pA, which is marked by the solid (blue) line. For $t_{BTO} \leq 5$ uc, $I_{ON}$ and $I_{OFF}$ become higher than the dotted curves due to the large leakage current. For $t_{BTO} = 10$ uc, the $I_{OFF}$ value (marked by the pink open square symbol) comes mainly from the noise, not from the true tunneling current. (b) $t_{BTO}$ dependent TER at -0.4 V. The TER calculated from Brinkerman's tunneling model is also plotted using a dotted line. For $t_{BTO} \leq 5$ uc, the degradation of TER is due to the large leakage current contributions to both $I_{ON}$ and $I_{OFF}$. For $t_{BTO} > 8$ uc, the degradation of TER is due to the detection limit of small $I_{OFF}$. These two TER degradations are marked by open and solid arrows, respectively. In (a) and (b), the vertical error bars correspond to standard deviations of current and TER values from 50 devices.

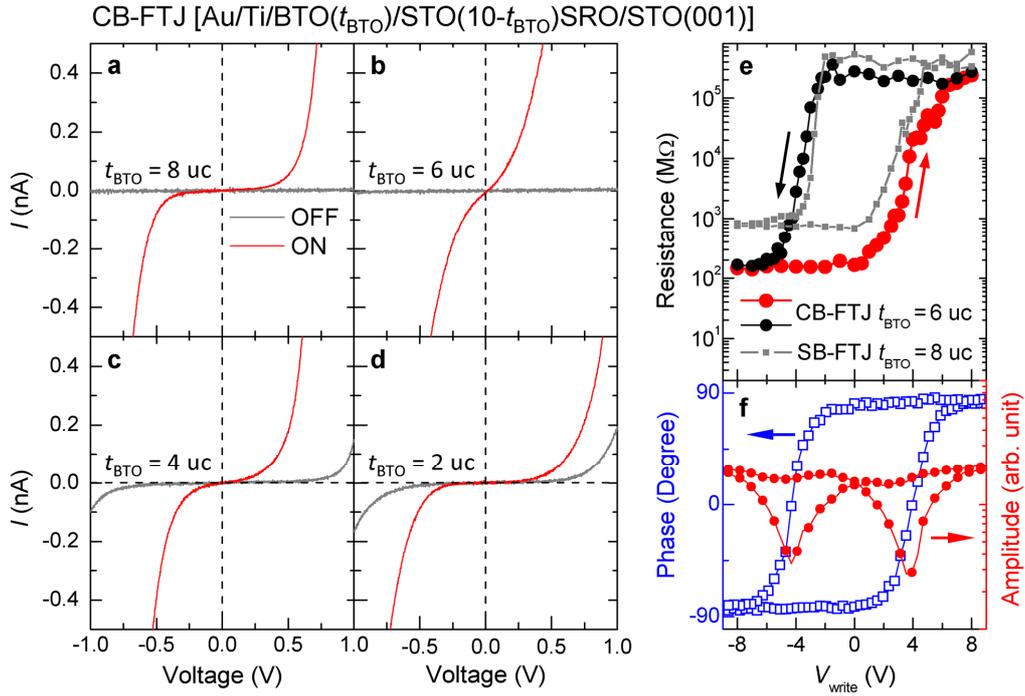

**Figure 5.** TER and resistive-switching behaviors in CB-FTJs. (a-d) *I-V* curves measured from the CB-FTJ with various $t_{BTO}$. (e) The *R-V*$_{write}$ curves with $V_{read}$ = -1 V after successive voltage pulses of 100 ms were applied. (f) PFM phase-voltage hysteresis loop and amplitude-voltage butterfly loop. (e) and (f) were measured from the CB-FTJ with $t_{BTO}$ = 6 uc, which clearly show that resistive-switching behavior occurs at the FE coercive fields. The *R-V*$_{write}$ curves of the best SB-FTJ ($t_{BTO}$ = 8 uc) are also shown in (e) for comparison.

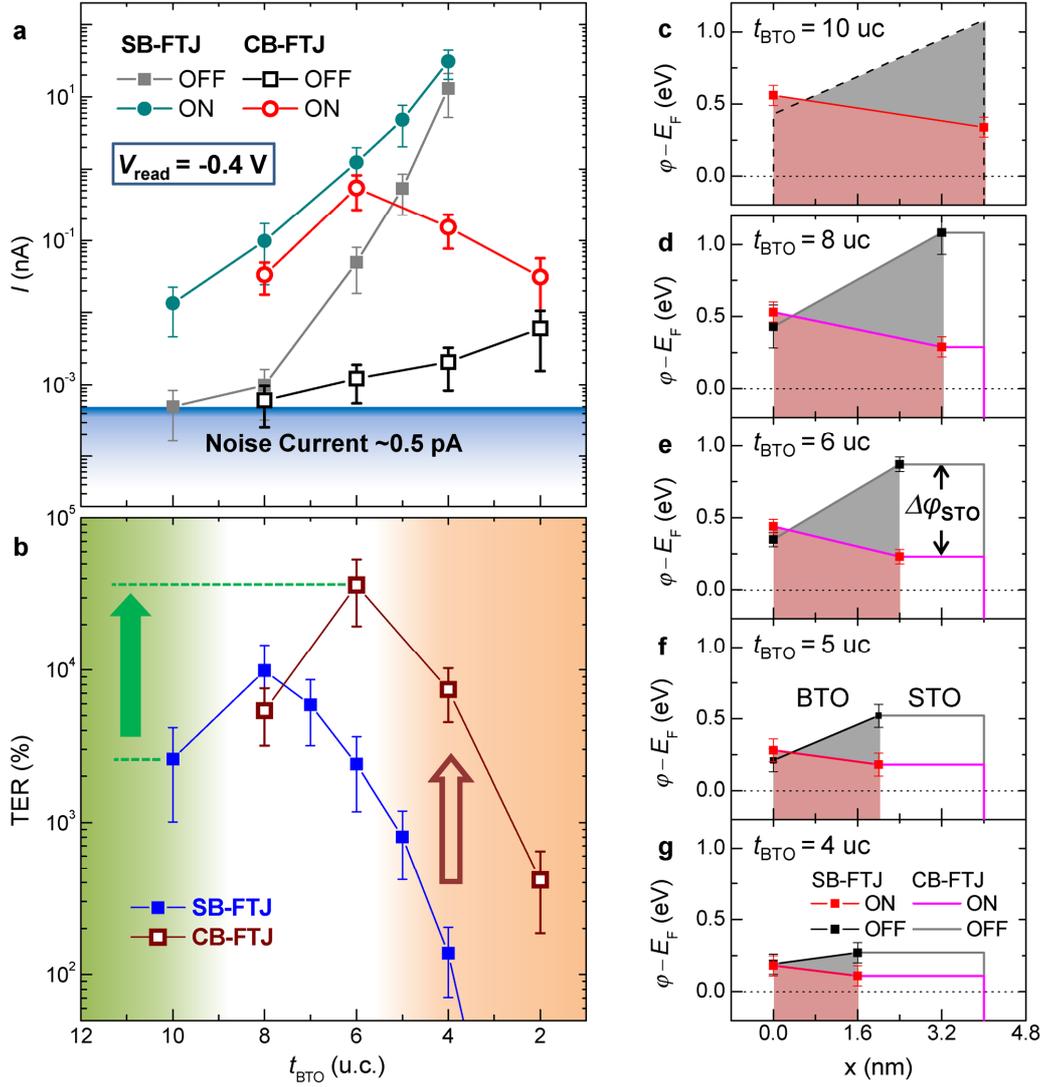

**Figure 6.** Barrier-composition-dependent device performances of CB-FTJs. (a,b) $t_{BTO}$-dependent (a) $I_{ON}$, $I_{OFF}$ and (b) TER values at -0.4 V, measured from the CB-FTJs with various $t_{BTO}$ ($t_{total} \equiv$ 10 uc). The vertical error bars correspond to standard deviations of current and TER values from 50 devices. The $I_{ON}$ and $I_{OFF}$ (TER values) of SB- FTJs are also included for comparison. (c-g) Barrier potential profiles of SB-FTJs and CB-FTJs with various $t_{BTO}$. The barrier height parameters of BTO were determined by fitting the *I-V* curves of the SB-FTJs. The BTO barrier potential profiles at ON (OFF) states are plotted using circles (squares), and highlighted by the shaded areas. Note that the barrier potential profile for $t_{BTO}$ = 10 uc at OFF state is plotted with a dashed line, because it cannot accurately be determined from $I_{OFF}$ under the noise current background. For the CB-FTJ, the barrier potential profiles are plotted in solid lines. We assumed that the barrier height parameters of BTO were the same as those of SB-FTJ with same $t_{BTO}$, and STO barrier heights were pinned with those of BTO at the STO/BTO interface. The Fermi levels ($E_F$) are marked by the dotted lines. The definition of $\Delta\varphi_{STO}$ is illustrated in (e).

REFERENCES

1. Waser, R.; Aono, M. *Nat. Mater.* 2007, *6* (11), 833-840.

2. Tsymbal, E.; Y. Gruverman, A.; Garcia, V.; Bibes, M.; Barthelemy, A. *MRS Bull.* **2012**, 37 (02), 138-143.

3. Garcia, V.; Bibes, M. *Nat. Commun.* **2014**, 5, 4289.

4. Esaki, L.; Laibowitz, R. B.; Stiles, P. J. *Tech. Discl. Bull.* **1971**, 13 (8), 2161.

5. Zhuravlev, M. Ye.; Sabirinaov, R. F.; Jaswal, S. S.; Tsymbal, E. Y. *Phys. Rev. Lett.* **2005**, 94 (24), 246802.

6. Velev, J. P.; Duan, C. -G.; Belashchenko, K. D.; Jaswal, S. S.; Tsymbal, E. Y. *Phys. Rev. Lett.* **2007**, 98 (13), 137201.

7. Kohlstedt, H.; Pertsev, N.; A. Contreras, J. R.; Waser, R. *Phys. Rev. B* **2005**, 72 (12), 125341.

8. Tsymbal, E. Y.; Kohlstedt, H. *Science* **2006**, 313 (5784), 181-183.

9. Maksymovych, P.; Jesse, S.; Yu, P.; Ramesh, R.; Baddorf, A. P.; Kalinin, S. V. *Science* **2009**, 324 (5933), 1421-1425.

10. Gruverman, A.; Wu, D.; Lu, H.; Wang, Y.; Jang, H. W.; Folkman, C. M.; Zhuravlev, M. Ye.; Felker, D.; Rzchowski, M.; Eom, C.-B.; Tsymbal, E. Y. *Nano Lett.* **2009**, 9 (10), 3539-3543.

11. Garcia, V. Fusil, S.; Bouzehouane, K.; Enouz-Vedrenne, S.; Mathur, N. D.; Barthelemy, A.; Bibes, M. *Nature* **2009**, 460 (7251), 81-84.

12. Pantel, D.; Goetze, S.; Hesse, D.; Alexe, M. *ACS Nano* **2011**, 5 (7), 6032-6038.

13. Chanthbouala, A.; Crassous, A.; Garcia, V.; Bouzehouane, K.; Fusil, S.; Moya, X.; Allibe, J.; Dlubak, B.; Grollier, J.; Xavier, S.; Deranlot, C.; Moshar, A.; Proksch, R.; Mathur, N. D.; Bibes M.; Barthelemy, A. *Nat. Nanotech.* **2012**, 7 (2), 101-104.

14. Wen, Z.; Li, C.; Wu, D.; Li, A.; Ming, N. *Nat. Mater.* **2013**, 12 (7), 617-621.

15. Hu, W. J.; Wang, Z.; Yu, W.; Wu, T. *Nat. Commun.* **2016**, 7, 10808.

16. Li, C.; Huang, L.; Li, T.; Lü, W.; Qiu, X.; Huang, Z.; Liu, Z.; Zeng, S.; Guo, R.; Zhao, Y.; Zeng, K.; Coey, M.; Chen, J.; Ariando; Venkatesan, T. *Nano Lett.* **2015**, 15 (4), 2568-2573.

17. Yin, Y. W. Burton, J. D.; Kim, Y. -M.; Borisevich, A. Y.; Pennycook, S. J.; Yang, S. M.; Noh, T. W.; Gruverman, A.; Li, X. G.; Tsymbal, E. Y.; Li, Q. *Nat. Mater.* **2013**, 12 (5), 397-402.

18. Jiang, L.; Choi, W. S.; Jeen, H.; Dong, S.; Kim, Y.; Han, M. -G.; Zhu, Y.; Kalinin, S. V.; Dagotto, E.; Egami, T.; Lee, H. N. *Nano Lett.* **2013**, 13 (12), 5837-5843.

19. Lu, H.; Lipatov, A.; Ryu, S.; Kim, D. J.; Lee, H.; Zhuravlev, M. Y.; Eom, C. B.; Tsymbal, E. Y.; Sinitskii, A.; Gruverman, A. *Nat. Commun.* **2014**, 5, 5518.

20. Boyn, S.; Garcia, V.; Fusil, S.; Carrétéro, C.; Garcia, K.; Xavier, S.; Collin, S.; Deranlot, C.; Bibes, M.; Barthélémy, A. *APL Mater.* **2015**, 3 (6), 061101.

21. Yamada, H.; Garcia, V.; Fusil, S.; Boyn, S.; Marinova, M.; Gloter, A.; Xavier, S.; Grollier, J.; Jacquet, E.; Carrétéro, C.; Deranlot, C.; Bibles, M.; Barthélémy, A.; *ACS Nano* **2013**, 7 (6), 5385-5390.

22. Sun, P.; Wu, Y. -Z.; Cai, T. -Y; Ju, S. *Appl. Phys. Lett.* **2011**, 99 (5), 052901.

23. Velev, J. P.; Duan, C. -G.; Burton, J. D.; Smogunov, A.; Niranjan, M. K.; Tosatti, E.; Jaswal, S. S.; Tsymbal, E. Y. *Nano Lett.* **2009**, 9 (1), 427-432.

24. Lu, H.; Liu, X.; Burton, J. D.; Bark, C. -W.; Wang, Y.; Zhang, Y.; Kim, D. J.; Stamm, A.; Lukashev, P.; Felker, D. A.; Folkman, C. M.; Gao, P.; Rzchowski, M. S.; Pan, X. Q.; Eom , C. -B.; Tsymbal, E. Y.; Gruverman, A. *Adv. Mater.* **2012**, 24 (9), 1209-1216.

25. Wu, Y. *J. Appl. Phys.* **2012**, 112 (5), 054104.

26. Kim, Y. S.; Kim, D. H.; Kim, J. D.; Chang, Y. J.; Noh, T. W.; Kong, J. H.; Char, K.; Park, Y. D.; Bu, S. D.; Yoon, J. -G.; Chung, J. -S. *Appl. Phys. Lett.* **2005**, 86 (10), 102907.

27. Zhuravlev, M. Ye.; Wang, Y.; Maekawa, S.; Tsymbal, E. Y. *Appl. Phys. Lett.* **2009**, 95 (5), 052902.

28. Caffrey, N. M.; Archer, T.; Rungger, I.; Sanvito, S. *Phys. Rev. Lett.* **2012**, 109 (22), 226803.

29. Ruan, J.; Qiu, X.; Yuan, Z.; Ji, D.; Wang, P.; Li, A.; and Wu, D. *Appl. Phys. Lett.* **2016**, 107 (23), 232902.

30. Choi, K. Biegalski, J. M.; Li, Y. L.; Sharan, A.; Schubert, J.; Uecker, R.; Reiche, P.; Chen, Y. B.; Pan, X. Q.; Gopalan, V.; Chen, L.-Q.; Schlom, D. G.; Eom, C. B. *Science* **2004,** 306 (5698), 1005-1009.

31. Lichtensteiger, C.; Triscone, J. -M.; Junquera, J.; Ghosez, P. *Phys. Rev. Lett.* **2005**, 94 (4), 047603.

32. Kim, D. J.; Lu, H. Ryu, S.; Bark, C. -W.; Eom, C. -B.; Tsymbal, E. Y.; Gruverman, A. *Nano Lett.* **2012**, 12 (11), 5697-5702.

33. Soni, R.; Petraru, A.; Meuffels, P.; Vavra, O.; Ziegler, M.; Kim, S. K.; Jeong, D. S.; Pertsev, N. A.; Kohlstedt, H. *Nat. Commun.* **2014**, 5, 5414.

34. Kohlstedt, H. Petraru, A.; Szot, K.; Rüdiger, A.; Meuffels, P.; Haselier, H.; Waser, R.; Nagarajan, V. *Appl. Phys. Lett.* **2008**, 92 (6), 062907.

35. Fujisawa, H.; Shimizu, M.; Niu, H.; Nonomura, H.; Honda, K. *Appl. Phys. Lett.* **2005**, 86 (1), 012903.

36. Griffiths, D. J. *Introduction to Quantum Mechanics* 2nd ed.; Prentice Hall: New York, 2005; pp 274-283.

37. Brinkman, W. F.; Dynes, R. C.; Rowell, J. M. *J. Appl. Phys.* **1970**, 41 (5), 1915-1921.

38. Guo, R. Wang, Z.; Zeng, S.; Han, K.; Huang, L.; Schlom, D. G.; Venkatesan, T.; Ariando; Chen, J. S. *Sci. Rep.* **2015**, 5, 12576.

39. Junquera, J.; Ghosez, P. *Nature*, **2003**, 422 (6931), 506-509.

40. Lichtensteiger, C.; Fernandez-Pena, S.; Weymann, C.; Zubko, P.; Triscone, J. -M. *Nano Lett.* **2014**, 14 (8), 4205-4211.